**Title: Investigating Epithelial-To-Mesenchymal Transition with Integrated Computational and Experimental Approaches**

**Jianhua Xing[1,2], Xiao-Jun Tian[3]**

[1]Department of Computational and Systems Biology, and UPMC-Hillman Cancer Center, School of Medicine, University of Pittsburgh, Pittsburgh, Pennsylvania, 15261, USA;

[2]UPMC-Hillman Cancer Center, University of Pittsburgh, Pittsburgh, PA, USA;

[3]School of Biological and Health Systems Engineering, Arizona State University, Tempe, AZ, 85287, USA

[*]To whom correspondence should be addressed:

Jianhua Xing, Ph.D., Department of Computational and Systems Biology, University of Pittsburgh, 3501 Fifth Ave, Pittsburgh, PA 15261, E-mail: xing1@pitt.edu.

Xiao-Jun Tian, Ph.D., School of Biological and Health Systems Engineering, Arizona State University, 781 S Terrace Rd, Tempe, AZ 85287, E-mail: xiaojun.tian@asu.edu;

**Abstract**

The transition between epithelial and mesenchymal (EMT) is a fundamental cellular process that plays critical roles in development, cancer metastasis, and tissue wound healing. EMT is not a binary process but involves multiple partial EMT states that give rise to a high degree of cell state plasticity. Here, we first reviewed several studies on theoretical predictions and experimental verification of these intermediate states, the role of partial EMT on kidney fibrosis development, and how quantitative signaling information controls cell commitment to partial or full EMT upon transient signals. Next, we summarized existing knowledge and open questions on the coupling between EMT and other biological processes, such as the cell cycle, epigenetic regulation, stemness, and apoptosis. Taken together, EMT is a model system that has attracted increasing interests for quantitative experimental and theoretical studies.

**Introduction**

Epithelial cells are adherent cells that form tightly adherent top layers of cavities, glands, and other structures throughout the body. In contrast, mesenchymal cells are loosely connected to each other and develop into connective tissues. Transition from the epithelial to the mesenchymal phenotype, known as epithelial-to-mesenchymal transition (EMT), is typically characterized by loss of cell-cell adhesion and apicobasal polarity and increased cell motility and invasion [1, 2]. EMT and the reverse process, mesenchymal-to-epithelial transition (MET), play important roles in embryonic development, tissue regeneration, and wound healing [3-5]. EMT also takes place in some pathological processes such as cancer progression and fibrosis formation in organs including the lungs, liver, and kidney. A plethora of molecular species regulate the EMT process at various levels and are potential pharmaceutical targets [6, 7]. Therefore, due to the biological importance and biomedical significance of EMT, in recent years there has been an explosive growth on the number of EMT-related studies in both physiological and pathological contexts.

Concurrently, EMT has also attracted increasing interests from the biological physics community. First, mathematical modeling can help unravel regulatory mechanisms underlying the seemingly complex EMT process. At the cellular level, the EMT process involves global reprogramming of cell transcriptome, epigenome, chromosome structure, and many phenotypic properties. Multiple molecular species form an interconnected intracellular network, and the functional roles of each one must be placed in the context of the overall regulatory network. At the tissue level, epithelial cells receive cues from diffusive extracellular factors and neighboring cells and make a choice on EMT or other cell fates. Combined with quantitative measurements, mathematical modeling has emerged as a powerful tool for studying the biology of EMT at different levels. Second, EMT as a biological process can also serve as a model system to address physics-oriented questions such as information transmission and transitions dynamics between multiple metastable states of a nonlinear dynamical system kept far from thermodynamic equilibrium.

In this review, we first discuss some studies that use integrated quantitative experiments and mathematical modeling to address certain EMT-related questions. We then provide our perspectives on a few problems that may attract attention in the near future and the interplay between the biology of EMT, nonlinear dynamics, and none-equilibrium physics. Due to the length limit, we will restrict our discussions to selected topics and put more focus on our own work. We apologize in advance for not mentioning a large number of studies, and the readers can find some related discussions from other reviews in this theme issue.

**1. Quantitative studies revealed a stepwise EMT dynamics through intermediate states.**

Cancer metastasis causes more than 90% of cancer-related deaths [8]. Therefore, it is a key strategy in cancer treatment to reduce the probability of metastasis, for which lack of mechanistic understanding of metastasis remains a major limiting factor. Cancer cells in primary epithelial solid tumors are epithelial cells that hold tightly with neighboring cells. To metastasize, epithelial cancer cells must break such intercellular connections and become more mobile. A proposal that received lots of attention is that cancer cells use EMT to convert tightly packed solid tumor epithelial cells into migratory mesenchymal cells; the latter enter into the circulatory system and become circulating tumor cells (CTCs). Some CTC cells survive the harsh environment of the circulatory system, and then manage to leave and transform back into epithelial cells through the MET process to serve as seeding cells for metastatic tumor formation [3-5, 9-18]. Results from various cancer types support the theory [5, 19-21].

Multiple lines of evidence, however, also challenge the EMT theory of metastasis. Some mouse studies reach a conclusion that EMT is not required for cancer metastasis but contributes to chemoresistance in pancreatic and lung cancer with different EMT phenotypes responding to treatments differently [22, 23]. There are heated debates on how to interpret these results [24, 25]. Close examination of the two studies reveals that while a complete EMT might not significantly contribute to metastasis, the two studies do not rule out incomplete (or partial) EMT as a candidate. EMT proceeds through several intermediate phenotypes collectively known as the partial EMT (pEMT) state (Fig. 1A) [26-30]. The pEMT state retains some characteristics of epithelium but also shows features of mesenchymal cells [31], and CTC cells typically move in clusters and possess the pEMT feature [10, 32].

It turns out the pEMT phenotype also appears in other contexts such as developmental processes [33, 34] and fibrosis discussed below. These experimental and clinical observations lead to some mechanistic questions: can pEMT exist as one or multiple stable phenotypes, and how does it originate? Several theoretical studies have been devoted to unraveling the mechanism(s) of generating the multiple EMT-related phenotypes, with a few of them detailed below.

TGF-β is one of the major inducers of EMT and plays essential roles in the developmental and normal physiological process. Dysregulation of TGF-β signal leads to complex diseases such as cancer, fibrosis, and Alzheimer's disease. Previous studies identified a core regulatory network for TGF-β-induced EMT (Fig. 1B) [35-37]. This network contains two double-negative feedback loops: one between transcription factor SNAIL1 and the miR-34 microRNA family, and another between transcription factors ZEB1/2 and the miR-200 family. Based on this network, Tian *et al*. proposed a Cascading Bistable Switches (CBS) mechanism [38]. Through bifurcation analysis, the model predicts that TGF-β induced EMT proceeds through two coupled reversible and irreversible bistable switches (Fig. 1C). The SNAIL1/miR-34 double-negative feedback loop controls the first bistable switch and the transition from the epithelial to the pEMT state. The transition is reversible, so when the exogenous TGF-β is removed, cells return to the epithelial state. Further increase of the exogenous TGF-β level flips the second bistable switch formed by the ZEB/miR-200 double-negative feedback loop, and cells change from the pEMT state to the fully mesenchymal state. This switch flipping also releases inhibition of miR-200 on endogenous TGF-β expression, which makes the second bistable switch irreversible. Subsequent experimental studies with human MCF10A cells [39] confirmed several key model predictions, and the flow cytometry data in Fig. 2 shows such an example. The E-cadherin/Vimentin two-color flow cytometry sorts cells into three groups corresponding to epithelial (E with E-cadherin$^{high}$Vimentin$^{low}$), partial EMT (P with E-cadherin$^{medium}$Vimentin$^{medium}$), and mesenchymal (M with E-cadherin$^{low}$Vimentin$^{high}$) states, respectively. These authors also demonstrated that cells maintain the partial EMT phenotype after weeks of culturing under treatment of intermediate TGF-β concentration. That is, the pEMT state is a stable phenotype rather than a transient intermediate. Cells also segregate into SNAIL1$^{low}$ and SNAIL1$^{high}$ sub-populations under intermediate TGF-β concentration (i.e., 0.5 and 1 ng/ml), reflecting the change of the first switch. At higher (1 and 2 ng/ml) TGF-β concentration, ZEB1$^{low}$ and ZEB1$^{high}$ sub-populations coexist, reflecting the change of the second switch. Therefore, there is a remarkable agreement between experimental results and model predictions. Tian *et al.* made some mathematical approximations on modeling the mRNA-miRNA interactions [38], which raised theoretical concerns on the existence of the SNAIL1-associated switch. In a revised model [39], the mRNA-miRNA mutual interactions and the dynamics of individual molecular species were explicitly modeled. A more thorough theoretical analysis further confirmed that a generic mRNA-miRNA mutual interaction motif can generate rich dynamical patterns including bistability [40]. That is, these two follow-up studies support the theoretical conclusion in the original study [38].

Based on essentially the same core network, Lu *et al.* proposed a different Ternary Chimera Switch (TCS) mechanism and predicted a different partial EMT state [41]. These researchers noticed that SNAIL1 acts as its own transcription inhibitor. Therefore, they suggested that the SNAIL1-miR-34 module is not a bistable switch but serves as a noise filter to prevent inadvertent EMT. Instead, they showed that the ZEB/miR-200 feedback loop coupled to a direct or indirect self-activation of ZEB form a toggle switch that can generate the three different EMT phenotypes.

It might seem surprising that qualitatively different mechanisms are proposed based on essentially the same network. The differences reflect that network dynamics are determined by both its topology and parameter values, the latter being more context-dependent. While the MCF10A data supports the CBS model, it is possible that the TCS model describes EMT dynamics of some other cell lines and cells in specific contexts, where SNAIL1 plays a less essential role on generating the pEMT state. From a more abstract level, the two models agree that both suggest partial EMT state(s) can arise from combinations of feedback network motifs. Moreover, these two models are not necessarily exclusive. Actually, the core regulatory network is embedded in a larger EMT network that permits more stable states, analogous to the rugged protein folding landscapes. Indeed, Hong *et al*. identified one additional partial EMT state by adding an OVOL2-ZEB positive feedback loop to the core network [42]. A role of OVOL2 in stabilizing the partial EMT state was also reported [43, 44]. Computational models for larger EMT networks have indeed highlighted the existence of multiple partial EMT states. For instance, Boolean models for EMT networks that treat a gene expression level in a discrete way to be either ON (1) or OFF (0) identified states co-expressing various epithelial and mesenchymal genes [45-47]. Increasing *in vitro* and *in vivo* evidence show multiple intermediate partial EMT states. Huang *et al*. analyzed 1,538 ovarian tumors and identified multiple partial EMT states [48-50]. Similar plasticity was observed in a panel of non-small cell lung cancer and ovarian cancer cell lines [51]. Six intermediate partial EMT states were found during cancer metastasis in a skin cancer mouse model *in vivo* [52].

Taken together, integrated experimental and theoretical efforts have provided critical insights on EMT regulation, especially on the nature and dynamical characteristics of the pEMT state. The reversibility of the pEMT state is likely necessary for development, resolution after wound healing, and cancer metastasis.

## 2. Model studies guided experiments to reveal double-edged roles of partial EMT on injury repair and fibrosis formation.

Aside from being observed in cancer metastasis, EMT can also be found in fibrosis progression. Fibrosis is formed with excessive deposition of fibrous connective tissues secreted by cells like myofibroblasts, often during the process of recovery from injury or fibrogenesis [53]. The latter refers to the process of gradual formation and progression of fibrotic patches that interfere with normal architecture and function of underlying tissues and organs. Progression of fibrosis is generally irreversible and eventually leads to organ failure. Currently, there is no effective treatment. For example, patients that have recovered from acute kidney injuries (AKI) have a much higher risk of later developing chronic kidney diseases (CKD) manifested with fibrosis, and eventually end-stage renal disease (ESRD) that requires dialysis or kidney transplantation [54].

It is a well-established phenomenon that EMT takes place during fibrogenesis of various organs such as the liver, lungs, and kidney, but the exact role of EMT was under heated debate for decades [55-57]. One proposal is that epithelial cells undergo EMT and contribute to the population of myofibroblasts in injured tissues. However, LeBleu *et al* used a lineage tracing approach and showed that transition of the tubular epithelial cell to myofibroblast contributes to just 5% of all the possible sources including resident fibroblasts proliferation, bone marrow and endothelial-to-mesenchymal transition program [53]. Later, two studies demonstrated that most tubular epithelial cells undergo pEMT instead of full EMT [58, 59]. Those tubular cells in the pEMT state stay in the original place and do not transit to myofibroblasts directly. Instead, they secrete multiple cytokines, such as TGF-β and WNT, and indirectly promote activation and proliferation of myofibroblasts. Both studies show that perturbation of pEMT, such as by knockout of SNAIL1 or another EMT regulator TWIST1, significantly recedes renal fibrosis induced in a mouse kidney injury model. Taken together, these two papers demonstrated that pEMT instead of full EMT contributes to the progression of renal fibrosis and may be a potential target for pharmaceutical intervention.

The observation that pEMT contributes detrimentally to fibrogenesis raises an intrigue question: why is the pEMT program activated during the process of repairing kidney injury? Tian *et al*. performed combined mathematical modeling and mouse ischemic reperfusion injury model studies and showed that pEMT cells serve an important role of relaying and amplifying signals to recruit injury repair cells and accelerate the repair process [60]. In normal conditions, fibroblasts exist in a very low density and sparsely reside within the narrow interstitial space between kidney functional units called nephrons [58]. Injured nephrons send extracellular signals such as WNT to activate and recruit fibroblasts and other immune cells. However, WNT molecules have a short half-life and mainly mediate close-range signaling [61-63]. Furthermore, WNT diffusion within the tortuous and narrow interstitial space is likely subdiffusive, as anomalous diffusion has been observed in similar narrow brain extracellular spaces [64-70]. Therefore, diffusion alone is not efficient to propagate the repair signals from an injury site to distant sites for recruiting injury-repair cells. Instead along the way of diffusion WNT molecules induce tubular epithelial cells to undergo pEMT, and the pEMT cells secrete their own WNT molecules and serve as repair "signal relay and amplification stations". This relay mechanism is necessary and critical for sending repair signals over long distances, possibly in the form of spatial trigger waves [71]. Disruption of this process slows down the repair process and increases the risk of death after AKI. After injuries are repaired, the inflammatory response needs to be resolved, so the pEMT cells convert back to epithelial cells, and myofibroblasts either undergo apoptosis or convert back to resident fibroblasts. Incomplete resolution characterized by residual pEMT cells leads to sustained myofibroblast activation and fibrosis development. Therefore, pEMT facilitates quick repair. However, excessive pEMT also increases the risk of fibrosis, and these two opposite effects were indeed observed in mouse model studies. Thus, the potential severity and lethality of AKI can explain why a repair system has evolved that exploits pEMT for augmented and accelerated repair, even at the expense of increased risk of chronic disease.

Realizing that pEMT activation is a double-edged sword also has profound medical implications. Effective biomedical intervention must take into account the nonlinear dynamic nature of a biological system, as opposed to simply modulating selected pharmaceutical targets monotonically. Tian *et al*. showed that a sequential application of WNT agonist and antagonist reduces rates of both mouse death and fibrosis development [60].

3. **Iterations between experiments and modeling uncovered a mechanism how cells interpret quantitative signaling information to decide on EMT commitment.**

As shown in the two above examples, multiple stimuli such as TGF-β, sonic Hedgehog (SHH), and WNT can induce EMT [4]. This existence of multiplex EMT inducing stimuli might seem unexpected since under most *in vivo* conditions normal epithelial cells need to maintain the phenotypic integrity despite stochastic fluctuations of environmental and intrinsic factors. An answer to this paradox resides in realization that cells read and respond to quantitative information of the stimuli and make a cell fate decision accordingly. That is, only specific signals (or combinations of signals) exceeding a certain threshold of strength and duration can induce EMT/partial EMT (Fig. 3A) [38]. Therefore, detection and transmission of the temporal quality of the stimulating signals is an important cellular mechanism in regulating the downstream effectors.

Following an iterative procedure between measurements and modeling, Zhang *et al*. revealed a composite network for TGF-β duration detection and regulation of SNAIL1 activation (Fig. 3B) [72]. The network can be divided into three modules: the canonical TGF-β-SMAD-SNAIL1 network as an early response module, a late response module formed by a GLI1-mediated positive feedback, and GSK3 that bridges the two modules. A short pulse of TGF-β activates SMAD but not GLI1 and generates only transient SNAIL1 expression; a pulse exceeding thresholds both in strength and duration or sustained TGF-β exposure activates both SMAD and GLI1 to generate two waves of SNAIL1 expression. Consequently, cells can respond differentially depending on TGF-β duration. Inhibition of GLI1 increases the duration threshold to activate the second wave of SNAIL1. Notice either pSMAD2/3 or GLI1 can distinguish only certain range of TGF-β duration. For example, the temporal profile of pSMAD2/3 may distinguish well between the absence of a TGF-β pulse and a short TGF-β pulse (e.g., 1-2 hours for human MCF10A cells), but may not be able to do so between short and long (e.g., > 2 hours for MCF10A cells) pulses; GLI1, on the other hand, can distinguish between short and long pulses. Then a combination of the two signaling proteins may encode the information for a broader range and lead to TGF-β duration-dependent SNAIL1 expression patterns. This scheme can be further extended with more readouts included.

In cell signaling there is an apparent paradox between the measured puzzling limited information content from the dynamic profile of one readout [73, 74] and the remarkable accuracy of cellular signaling. The work of Zhang *et al*. [72] suggests a possible solution would be for cells to use multiple readouts to form a temporally-ordered state space (TOSS) to encode the signal information. Consider a coarse-grained model for coding the TGF-β duration information with two readouts (pSMAD2/3 and GLI1). Assume that each readout can exist in two distinct states (0 and 1) and show statistically uncorrelated dynamics at two (coarse-grained) time points. Theoretically, the two readouts form a TOSS with 16 as the maximum number of states and can carry a maximum information capacity of four bits. Therefore, cells can interpret and transmit high content information, analogous to a computer chip using combinations of multiple binary gates.

As for EMT studies, SNAIL1 is highly pleiotropic and is a key regulator of EMT [75], so the detected rich temporal dynamics of Snail1 by Zhang *et al*. also shed light on understanding how cells make a decision to commit to EMT. The uncovered network serves as a temporal checkpoint that prevents spurious SNAIL1 activation and subsequent major cellular fate changes by transient fluctuations of EMT inducing factors. Under pathological conditions, a reduced threshold for activating sustained SNAIL1 expression likely contributes to a tilt of balance towards cell survival through EMT instead of apoptosis [76].

## 4. Selected topics for future quantitative studies

In a larger context, the EMT program is coupled to the networks regulating many other biological processes such as cell cycle and stemness. It remains an active research area on examining how these programs are coupled to and interfere with EMT, and how a cell makes a fate choice. There are already many extensive reviews focusing on tumorigenesis, cancer metastasis [77-82], drug resistance [83], and crosstalk with other signaling pathways including SHH and WNT [84]. We anticipate that quantitative approaches with combined computational modeling and experimental measurements will continue to complement other more traditional approaches and provide important insights to the regulation and physiological roles of EMT under different contexts. Given a potentially long list of problems and many existing reviews, below we will focus on a few selected topics that have been less emphasized previously.

### 4.1. EMT and Cell Cycle Arrest

Cells transiently enter cell cycle arrest during the course of EMT, but details on how the two processes are coupled remain unclear. TGF-β1--induced EMT is directly associated with P21-mediated G2 arrest in tubular epithelial cells [85]. Both TWIST1 and SNAIL1 induce P21-mediated G2 arrest in renal fibrosis [58]. Interestingly, in mouse AML-12 cells TGF-β1 induces

EMT only at the G1/S phase, while it induces apoptosis at the G2/M phase [86]. It is also reported that TGF-β1 and TGF-β3 play separate roles in cell cycle arrest and EMT [87]. Following cell cycle arrest induced by TGF-β1, TGF-β3 executes EMT and apoptosis [87]. Thus, it has been suggested that cells cease cell cycle as a prerequisite to facilitate EMT. Consistent with this requirement of cell cycle arrest, EMT is often associated with low proliferation but high invasive ability. Cells undergoing EMT also acquire resistance to two safeguard mechanisms: senescence and apoptosis [3].

Existing studies suggest that EMT and the cell cycle are coupled at various stages and can be cell-type specific. Here we summarize two hypothetical coupling schemes (Fig. 4). In both schemes, cells first have an increased concentration of P21/SNAIL1 and undergo cell cycle arrest and partial EMT (state Pα). The depth of the partial EMT state Pα depends on the level of P21, analogous to the depth control of cellular quiescence by an Rb-E2F switch [88], and sufficiently high concentrations of P21 may drive cells to the largely irreversible senescent state. The main difference between the two schemes is whether cells exit cell cycle arrest before (scheme 1 with another pEMT state Pβ), together with (scheme 2), or even after completion of EMT. Mechanistically, various factors regulate EMT and the cell cycle differently. Specifically, SNAIL1 has been shown to mainly initialize EMT and up-regulate P21, while TWIST1 is activated at a later stage to maintain EMT but represses P21 expression [89-91]. Different strengths of these interactions may result in either scheme 1 or 2, and quantitative measurements such as time-lapse single cell imaging are needed to distinguish these competing mechanisms.

### 4.2. EMT and Apoptosis

It is intuitive to think that EMT and other cell fates such as apoptosis are mutually exclusive. EMT is considered as a mechanism for cancer cells to escape from and survive through stressful conditions [92, 93]. This understanding is complicated by a recent study reporting an unexpected observation that in pancreatic ductal adenocarcinoma (PDA) cells EMT precedes apoptosis [94]. Both TGF-β induced EMT and SOX4 expression are generally considered pro-tumorigenic. SOX4 requires a transcription factor KLF5 as a cofactor to be anti-apoptotic and is pro-apoptotic without KLF5. However, in TGF-β sensitive cells, the TGF-β-SMAD-SNAIL pathway induced EMT represses KLF5. Therefore, crosstalk between the two presumably pro-tumorigenic pathways results in SOX4-induced lethal apoptosis. This surprising observation exemplifies the complexity and context-dependence of cell fate decision programs.

Another intriguing EMT related phenomenon is called anastasis, as opposed to apoptosis. Under stress a cell first appears to undergo apoptosis with apparent morphological change, then survives and switches to EMT instead. This process has been observed in multiple cell types and has become an active research area recently [95-97]. Figure 5 shows three human renal epithelial HK2 cells undergoing significant morphology change under TGF-β treatment. In the top, a cell underwent partial and then full EMT. It changed step-by-step from a regular polygon shape into a larger and spindle-shaped cell, characteristic of EMT. In the middle, a cell underwent apoptosis with membrane blebbing. The bottom shows an example of anastasis followed by EMT. This cell first showed membrane blebs, indicating that it was undergoing apoptosis, but then the cell managed to recover from apoptosis and undergo EMT instead. To understand the molecular mechanism of anastasis, Sun *et al* performed a transcriptome analysis of untreated, apoptotic and recovering human HeLa cells [96]. Their data challenged the traditional view that caspase activation is a no-return point for apoptosis. Instead, they found that mRNAs of some genes including *snail,* which are required for recovery from apoptosis, are transcribed but not translated when caspase is still activated. That is, cells are poised for recovery even when the apoptosis program is under execution, so they can recover quickly and undergo EMT if the apoptosis stress is removed. The observation that the process of apoptosis remains reversible even at a very late stage of the process raises a question that requires a quantitative answer: when is the no-return point of apoptosis? Computational modeling in combination with quantitative measurements will be useful in answering such a question.

**4.3. EMT and Stemness**

As discussed in section 1, the EMT regulatory program couples to a number of other cellular programs with shared regulators and crosstalk. Specifically, the possible relation between EMT and stemness in cancer cells has raised lots of attention [98-101]. The cancer stem cell (CSC) theory is a heatedly debated hypothesis on cancer heterogeneity and drug resistance [99, 101-106]. Originally the theory states that a fraction of cancer cells have the capacity of self-renewal and differentiation into other cancer cell types to sustain the cancer cell population, much like normal stem cells. Recent studies suggest that unlike normal stem cells, CSCs may come from dedifferentiation of non-CSCs [83, 107-109]. In an influential study, Weinberg and coworkers showed that induction of EMT in immortalized human mammary epithelial cells generates cells with stem cell properties [110]. Similarly, Morel *et al*. showed that activation of the RAS-MAPK pathway induces stemness characteristics in human mammary epithelial cells [111]. The two studies led to an explosive number of studies on the relation between EMT and CSC generation [99]. Notably Jolly *et al*. analyzed a combined EMT/stemness gene regulatory circuit and suggested that one can fine-tune the positioning of the stemness window along the EMT axis [79].

Here we propose a general strategy to study the coupling between EMT and stemness phenomenologically using breast cancer cells as an example system. First consider the EMT and stemness regulation separately. While the EMT axis is likely a continuum, for simplicity of discussion we consider three coarse-grained EMT phenotypes, E, P, and M. In practice two surface markers, CD44 and CD24, are widely used in breast cancer studies. Similarly, we consider two coarse-grained stem states, cancer stem-like cells (CD44$^{high}$CD24$^{low}$, denoted as "S" in the following discussions) and non-stem-like cells (CD44$^{low}$CD24$^{high}$, denoted as "N") [110]. If the EMT and CSC programs are weakly coupled, each program maintains the identity of its own state, and is only weakly perturbed by cross-talking with another one. Then one expects a complete state network with a total of six possible states with all possible transitions among them (Fig. 6A). On the contrary, if the two programs are strongly coupled, the number of populated states may be less since the EMT state of one cell may determine its stem state and vice versa. Notice that there are three EMT states but only two CSC states that have been identified. Then the relation between the "P" state and CSC states is unclear. Figure 6B &C show two possibilities and plausible pathways for the transitions, with the P state coupling to the N state in one pathway, and to the S state in another one. Potentially, another intermediate state CD44$^{medium}$CD24$^{medium}$, may exist, which would further complicate this roadmap. Establishing such a map of the EMT-CSC state network is of great importance not only for understanding cell phenotype transition, but also for developing effective cancer treatment strategies. Experimentally, one can distinguish the different schemes through multi-color flow cytometry with both EMT and stemness markers. More generally one can represent the coupled process with a continuous two-dimensional surface composed with the EMT and stemness axes. The proposed framework can also be applied to other processes coupled to EMT. From a theoretical perspective, the framework is analogous to representing an interacting many-particle system with a single-particle diabatic representation in quantum chemistry.

### 4.4. EMT and epigenetics

Epigenetic regulation is also an essential part of EMT regulation [14]. McDonald *et al*. examined TGF-β induced EMT of AML12 mouse hepatocytes [112]. During EMT, the global level of DNA methylation was barely changed, yet H3K9Me3 was reduced, and H3K4Me3/H3K36Me3 was increased on the genome-scale [112]. In another study on primary human mammary epithelial cells, Dumont *et al*. showed that epigenetic change during EMT has a determined temporal order, with inhibition of E-Cadherin transcription preceding methylation of its promoters. Furthermore, EMT transcription factors actively participate in the process of epigenetic regulation. For example, TWIST inhibits E-cadherin transcription by interacting with the deacetylase (Mi2/NuRD) complex, MTA2, RbAp46, Mi2 and HDAC2 [113]. SNAIL1 is able to recruit histone demethylase LSD1 to remove the dimethylation on H3K4me2 and thus promotes further repression of its targets including E-cadherin [114]. SNAIL1 can cause bivalent histone modifications of target gene promoters, which have both active marker H3K4me3 and repressive marker H3K27me3 [115]. Specifically, the bivalent epigenetic state was also found at the ZEB1 promoter, which enables breast cancer cells to be more plastic and enhances tumorigenicity [116]. It is noted that the bivalent epigenetic state is a poised state that renders a gene susceptible to reactivation [116]. We suggest that this is a possible feature of the pEMT state, which is reversible in both phenotypic level and epigenetic level. Taken together, integrated regulation at different layers greatly enhances both stability and plasticity of EMT phenotypes. This coupling is supported by the observed correlation among expression, gene spatial proximity, and histone modification patterns of genes regulated by common transcription factors in TGF-β treated MCF10A cells [117].

Physics-based models have been used to study how the coupling of epigenetic and genetic regulation controls cell fates [118-123]. Previously, Zhang *et al*. constructed a Potts-type model for epigenetic histone modification and showed that lateral interaction between nearest-neighbor enzymes is sufficient for collective epigenetic states [119]. Then by applying the epigenetic modeling framework to the differentiation process of olfactory sensory neurons, a theoretical model study by Tian *et al*. shed light on an intriguing puzzle about the 'one-neuron, one-receptor' phenomenon [118]. Their multiscale model by coupling the genetic and epigenetic regulation demonstrated how the physical principle of cooperativity provides a mechanistic explanation of this challenging biological problem. In future work, we expect to see more such model studies on describing the dynamics aspect of the coupling between epigenetic modifications and other mechanisms of gene regulation such as chromosome structure and transcription factors on regulating the dynamics and spectrum of EMT.

4.5. **Quantitative single cell techniques emerge as important tools for EMT studies**

Various methods and techniques have been introduced to address the above discussed and other EMT-related problems. Among them, single cell techniques emerge as powerful tools. Many intrinsic and extrinsic factors, such as cell density, cell cycle stage, and fluctuations of intracellular EMT regulator concentrations, may affect whether a cell commits to EMT and how fast that conversion process proceeds. Single cell studies are informative on dissecting the effects of these factors. We have discussed several studies using fixed cell-based techniques such as flow cytometry and fluorescence staining [39, 42, 72, 124, 125]. Single cell transcriptome studies have also been applied for understanding the roles of EMT in organogenesis and disease development [126-129]. Using live cell imaging at single cell resolution, Kuo and Krasnow recorded a slithering mode of cell migration during formation of ling neuroepithelial bodies, where cells transiently and partially lose epithelial characters and migrate to a next-neighboring site without completely detaching from the epithelium membrane throughout the process [130]. Using combined fixed and live cell imaging, Burute *et al*. observed centrosome repositioning during the process of EMT [131]. Compared to fixed cells, live cell studies provide information on the temporal correlation of intracellular dynamics for individual cells but are technically more challenging. Below we will review a few recent technical advances that are expected to facilitate single cell imaging studies on EMT.

Quantitative single cell image analysis starts with single cell segmentation to identify and outline regions of interest in an image, which is necessary for subsequently extracting quantitative single-cell information [132-134]. Accurate segmentation is also a prerequisite for cell tracking since errors of cell segmentation propagate along single-cell trajectories. For EMT studies tightly packed epithelial cells further increase the difficulty of correct cell segmentation. Several commonly used tools ImageJ [135] and CellProfiler [136] typically require time-consuming and labor-intensive fine-tuning of multiple parameters manually, which quickly becomes impractical with a large amount of time-lapse images. Machine learning based automated image analysis methods have become alternative choices in recent years [137]. Van Valen *et al*. showed that for segmenting live cell images deep convolutional neural networks (DCNN) significantly outperform other conventional methods both in accuracy and required efforts [138]. Wang *et al*. developed a method that combines the strength of DCNN and the conventional watershed algorithm, which segment densely packed epithelial cells with improved accuracy compared with direct application of DCNN [139]. Eschweiler *et al*. also showed that preprocessing 3D confocal images with CNN improves the accuracy of watershed-based segmentation [140].

For fluorescence-based imaging, generating cell lines with fluorescence labeling is another challenge, and the advent of CRISPR-based techniques greatly accelerate such processes. CRISPR-based endogenous knock-in labeling is desirable for quantitative biology studies since the "tagged" proteins are not overexpressed, and cellular dynamics are thus less perturbed. CRISPR-based gene knock-in can also facilitate generating model systems such as a fusion gene for cancer development. Knock-in is technically more challenging than knock-out. One possibly unexpected challenge is generating donor DNAs that contain the specific knock-in sequence and repair templates. Chen *et al*. developed an efficient and modular assembly procedure, which also allows convenient fusion of single guide RNAs (sgRNAs) to the constructs to achieve additional enhanced homology-directed repair efficiency [141]. Li *et al*. also developed an optimized procedure for synthesizing single-stranded DNA donors for CRISPR knock-in [142].

Chen *et al*. illustrate their procedure specifically for inserting fluorescence protein sequences into targeted genes [141]. The procedure can also be applied to generate cell lines for mRNA tagging. In this method, one needs to insert multiple copies of stem-loop sequences that can recruit RNA binding proteins such as MS2 fused with fluorescence proteins [143]. Difficulties of generating the cell lines have greatly limited its use in previous studies, and the advent of CRISPR techniques could drastically change the situation.

EMT is accompanied with a large chromosomal reorganization. Tracking the process can provide useful insight into how cells maintain phenotypic stability and plasticity. Besides fixed cell-based techniques such as Hi-C and FISH, CRISPR has been repurposed for labeling specific genomic loci in live cells with catalytically dead Cas9 (dCas9) fused with fluorescence proteins, or by extending sgRNAs with stem-loops that can recruit RNA binding proteins fused with fluorescence proteins [144-158]. The Xing lab is currently using this technique to study chromosome dynamics during EMT.

The CRISPR-Cas9 system has also been repurposed to regulate gene expression directly or edit epigenetic histone modifications and DNA methylation [159-161]. In both cases, sgRNAs recruit dCas9 fused with certain effectors to specific genomic locations. These techniques may be used for reversing the EMT process. One such target is the miR-200 family, where repressive histone marks and DNA methylation inactivate expression of members of this pro-epithelial miRNA family in many invasive cancer types showing mesenchymal and stem-like features [162, 163]. These techniques permit expression control of selective EMT related genes and can combine with other approaches such as optogenetics to achieve precise spatiotemporal control for single-cell studies.

5. **Discussion and concluding remarks**

Quantitative studies on EMT, both from experimental and theoretical perspectives, have only emerged recently, which already make it unfeasible to provide a thorough review here. We expect to see a rapid growth of such studies with new approaches and concepts from physics. Below we provide some perspective based on our own experiences.

The network is dynamic and cell type-specific. It is a standard practice to reconstruct biological networks from results obtained with different cell types and different stages of a dynamical process. The information is often static and incomplete. Therefore, caution needs to be taken on performing computational analysis of these networks.

In physics, one seeks general principles, which is challenged by diversity and heterogeneity of biological systems. It is unlikely to find a single model that can fit all the contexts. General principles, however, still exist at more abstract levels. For example, EMT studies demonstrate that coupled positive/negative feedback motifs are key network structures for regulating cell phenotypes [38, 41, 42].

Iteration between mathematical modeling and quantitative measurements is fruitful on advancing our understanding of EMT dynamics. Within a foreseeable future, it is unlikely that one can perform first principle type studies for complex biological systems such as EMT. Instead, a close and iterative interplay between modeling and experiments will remain as a necessary main theme. In their studies, Zhang *et al*. [72] first examined a simple mechanism based on available knowledge of TGF-β signaling, which states that the TGF-β induced pulsating SMAD2/3 activation initializes SNAIL1 expression, which is then self-maintained by the SNAIL1-miR-34 double-negative feedback loop. The mechanism is both biologically plausible and theoretically attractive but is insufficient to explain their quantitative time course data. The discrepancy led to their subsequent computational and experimental investigation that unravels the three-module network discussed in section 1. Notably following this iterative procedure the researchers discovered a new phenomenon of TGF-β signaling [72]. It is well documented that TGF-β leads to conversion of the GSK3 family proteins to a serine (S21 in GSK-3α and S9 in GSK-3β) phosphorylation form that is enzymatically inactive. Through computational analysis, these researchers found that this canonical GSK3 pathway could not explain a multi-phasic dynamics of GLI1, which is a downstream substrate of GSK3. Instead, immunostaining studies revealed that TGF-β first induces accumulation in the Golgi apparatus and endoplasmic reticulum of GSK3 in a much less studied tyrosine (Y279 in GSK-3α and Y216 in GSK-3β) phosphorylation form with increased enzymatic activity. This transient localization change has surprisingly escaped notice despite decades of extensive studies on TGF-β signal transduction. It is possibly due to the fact that this active form is only ~10% of the total GSK3 molecules and this percentage does not change significantly throughout TGF-β treatment, rather the change is mainly their cytoplasmic location. Zhang *et al*. further demonstrated that this increased local concentration of GSK3 active form has subtle but important functions on accelerating GLI1 accumulation upon TGF-β stimulation [72]. Further studies can examine whether spatial accumulation of the tyrosine phosphorylation form of GSK3 involves phase separation [164], and if so how the GSK3 solubility in different phases is modulated upon TGF-β treatment. Therefore, this discovery of new biology and subsequent emergence of new questions demonstrate the power of the iterative approach between modeling and quantitative measurements.

Similarly, constructive scientific debates also take place among modelers. In their original work, Tian *et al*. used a phenomenological form to model microRNA-mRNA mutual regulations [38]. Lu *et al*. pointed out the importance of treating these interactions more explicitly [41]. The latter prompted a further theoretical analysis that revealed surprisingly rich dynamics just with a simple microRNA-mRNA mutual regulation motif [40].

Studies focusing on a minimal abstract network and an extended detailed network are complementary. At the technical level, there are two general classes of approaches of studying a regulatory network. One is to start from a network that includes as many relevant molecular species as possible. The resultant hope-to-be-exhaustive network is large and can provide a direct comparison with many available experiments and make specific predictions. It is also straightforward to incorporate high throughput screening data such as microarray and RNA-seq into the model. The downside is that the number of adjustable parameters of a corresponding mathematical model heavily out-numbers possible constraints from feasibly available experimental data. Consequently, it is difficult to perform an inclusive analysis of the model. One approach is to sacrifice some quantitative features, for example, by performing Boolean type modeling or further network reduction. Another one is to aim at identifying a minimal network that sufficiently and quantitatively describes the dynamics of a process. The network is typically much smaller and is amenable for thorough analysis. The corresponding model has limited power to predict elements not explicitly considered in the core network. On the other hand, the small size of the network makes it feasible to be integrated with data-intensive quantitative measurements on the involved molecular species, and the interplay between modeling and quantitative data allows an expansion process from the core network to achieve one or a set of minimal and inclusive models.

Therefore, both approaches are complementary, each having their strengths and limitations. Both have been successfully used to address various questions, specifically in analyzing the EMT regulatory networks [38, 39, 41, 165]. An extended network can provide global dynamics with less resolution on the temporal evolution of individual constitute elements, and a minimal network can provide a quantitative description of a small number of network elements. For a specific problem, the choice of which approach to use should depend on the objective of the study and availability of data.

The above issue relates to some general theoretical challenges especially for systems out of thermodynamic equilibrium, such as how to reconstruct system dynamics from experimental data with a limited number of observables, and how to describe the dynamics of a primary system strongly coupled to a larger "environment". Xing and Kim derived a generalized Langevin equation using the Zwanzig-Mori projection formalism for systems without detailed balance [166]. In this formalism effects of the possibly infinite number of environmental degrees of freedom on the primary system are parameterized by a memory kernel and in general colored noise, similar to the case that solvent effects on a Brownian particle are modeled by a drag coefficient and white noises. Hilfinger *et al*. analyzed constraints that unspecified network components impose on fluctuations of the "primary" components being observed [167, 168]. The analysis separates properties of fluctuations due to local interactions within a primary system from influences from the coupled environment. Further theoretical development and applications will be desirable.

As concluding remarks, EMT is a fascinating biological process that physicists can provide unique insights on questions raised by biologists. Furthermore, it can also serve as a model system for which physicists can define physics-oriented questions such as transition dynamics and paths between nonequilibrium attractors.

**Acknowledgment**

This work was partially supported by the National Science Foundation [DMS-1462049], National Cancer Institute [R01CA232209], National Institute of Diabetes and Digestive and Kidney Diseases (R01DK119232), the Charles E Kaufman Foundation (KA2018-98550) to JX, and the startup fund from Arizona State University, School of Biological and Health Systems Engineering to XJT. We thank Drs Herbert Levine and Mohit Jolly for reading the manuscript and many helpful discussions. We thank Hanah Goetz for editing the manuscript. We apologize to the authors whose work could not be cited due to space.

**Reference:**

1. Kalluri, R. and R.A. Weinberg, *The basics of epithelial-mesenchymal transition.* J Clin Invest, 2009. **119**(6): p. 1420-8.
2. Radisky, D.C., *Epithelial-mesenchymal transition.* Journal of Cell Science, 2005. **118**(19): p. 4325-4326.
3. Thiery, J.P., et al., *Epithelial-mesenchymal transitions in development and disease.* Cell, 2009. **139**(5): p. 871-90.
4. Nieto, M.A., *The ins and outs of the epithelial to mesenchymal transition in health and disease.* Annu Rev Cell Dev Biol, 2011. **27**: p. 347-76.
5. Yang, J. and R.A. Weinberg, *Epithelial-mesenchymal transition: at the crossroads of development and tumor metastasis.* Dev Cell, 2008. **14**(6): p. 818-29.
6. Lee, J.M., et al., *The epithelial-mesenchymal transition: new insights in signaling, development, and disease.* J Cell Biol, 2006. **172**(7): p. 973-81.
7. Craene, B.D. and G. Berx, *Regulatory networks defining EMT during cancer initiation and progression.* Nature Reviews Cancer, 2013. **13**: p. 97.
8. Gupta, G.P. and J. Massagué, *Cancer Metastasis: Building a Framework.* Cell, 2006. **127**(4): p. 679-695.
9. Wynn, T.A. and T.R. Ramalingam, *Mechanisms of fibrosis: therapeutic translation for fibrotic disease.* Nat Med, 2012. **18**(7): p. 1028-40.
10. Yu, M., et al., *Circulating breast tumor cells exhibit dynamic changes in epithelial and mesenchymal composition.* Science, 2013. **339**(6119): p. 580-4.
11. Bonnomet, A., et al., *A dynamic in vivo model of epithelial-to-mesenchymal transitions in circulating tumor cells and metastases of breast cancer.* Oncogene, 2012. **31**(33): p. 3741-3753.
12. Chen, C.L., et al., *Single-cell analysis of circulating tumor cells identifies cumulative expression patterns of EMT-related genes in metastatic prostate cancer.* Prostate, 2013. **73**(8): p. 813-26.
13. Han, J.W. and Y. Yoon, *Epigenetic landscape of pluripotent stem cells.* Antioxidants & redox signaling, 2012. **17**(2): p. 205-223.
14. Tam, W.L. and R.A. Weinberg, *The epigenetics of epithelial-mesenchymal plasticity in cancer.* Nat Med, 2013. **19**(11): p. 1438-49.
15. Scheel, C. and R.A. Weinberg, *Cancer stem cells and epithelial–mesenchymal transition: Concepts and molecular links.* Seminars in Cancer Biology, 2012. **22**(5–6): p. 396-403.
16. Guarino, M., A. Tosoni, and M. Nebuloni, *Direct contribution of epithelium to organ fibrosis: epithelial-mesenchymal transition.* Human Pathology, 2009. **40**(10): p. 1365-1376.
17. Nieto, M.A. and A. Cano, *The epithelial-mesenchymal transition under control: global programs to regulate epithelial plasticity.* Semin Cancer Biol, 2012. **22**(5-6): p. 361-8.
18. Nieto, M.A., *Epithelial plasticity: a common theme in embryonic and cancer cells.* Science, 2013. **342**(6159): p. 1234850.
19. Garber, K., *Epithelial-to-mesenchymal transition is important to metastasis, but questions remain.* J Natl Cancer Inst, 2008. **100**(4): p. 232-3.
20. Tarin, D., E.W. Thompson, and D.F. Newgreen, *The fallacy of epithelial mesenchymal transition in neoplasia.* Cancer Res, 2005. **65**(14): p. 5996-6000.
21. Thompson, E.W., D.F. Newgreen, and D. Tarin, *Carcinoma invasion and metastasis: a role for epithelial-mesenchymal transition?* Cancer Res, 2005. **65**(14): p. 5991-5.
22. Fischer, K.R., et al., *Epithelial-to-mesenchymal transition is not required for lung metastasis but contributes to chemoresistance.* Nature, 2015. **527**: p. 472-476.
23. Zheng, X., et al., *Epithelial-to-mesenchymal transition is dispensable for metastasis but induces chemoresistance in pancreatic cancer.* Nature, 2015. **527**: p. 525-530.

24. Aiello, N.M., et al., *Upholding a role for EMT in pancreatic cancer metastasis.* Nature, 2017. **547**(7661): p. E7.
25. Brabletz, T., et al., *EMT in cancer.* Nat Rev Cancer, 2018. **18**(2): p. 128-134.
26. Leroy, P. and K.E. Mostov, *Slug is required for cell survival during partial epithelial-mesenchymal transition of HGF-induced tubulogenesis.* Mol Biol Cell, 2007. **18**(5): p. 1943-52.
27. Revenu, C. and D. Gilmour, *EMT 2.0: shaping epithelia through collective migration.* Curr Opin Genet Dev, 2009. **19**(4): p. 338-42.
28. Klymkowsky, M.W. and P. Savagner, *Epithelial-mesenchymal transition: a cancer researcher's conceptual friend and foe.* Am J Pathol, 2009. **174**(5): p. 1588-93.
29. Thomson, S., et al., *A systems view of epithelial-mesenchymal transition signaling states.* Clin Exp Metastasis, 2011. **28**(2): p. 137-55.
30. Futterman, M.A., A.J. Garcia, and E.A. Zamir, *Evidence for partial epithelial-to-mesenchymal transition (pEMT) and recruitment of motile blastoderm edge cells during avian epiboly.* Dev Dyn, 2011. **240**(6): p. 1502-11.
31. de Herreros, A.G., et al., *Snail family regulation and epithelial mesenchymal transitions in breast cancer progression.* J Mammary Gland Biol Neoplasia, 2010. **15**(2): p. 135-47.
32. Cheung, K.J., et al., *Polyclonal breast cancer metastases arise from collective dissemination of keratin 14-expressing tumor cell clusters.* Proceedings of the National Academy of Sciences, 2016.
33. Theveneau, E. and R. Mayor, *Neural crest delamination and migration: from epithelium-to-mesenchyme transition to collective cell migration.* Dev Biol, 2012. **366**(1): p. 34-54.
34. Duband, J.-L., *Diversity in the molecular and cellular strategies of epithelium-to-mesenchyme transitions: Insights from the neural crest.* Cell Adhesion & Migration, 2010. **4**(3): p. 458-482.
35. Siemens, H., et al., *miR-34 and SNAIL form a double-negative feedback loop to regulate epithelial-mesenchymal transitions.* Cell Cycle, 2011. **10**(24): p. 4256-4271.
36. Gregory, P.A., et al., *An autocrine TGF-beta/ZEB/miR-200 signaling network regulates establishment and maintenance of epithelial-mesenchymal transition.* Mol Biol Cell, 2011. **22**(10): p. 1686-98.
37. Bracken, C.P., et al., *A double-negative feedback loop between ZEB1-SIP1 and the microRNA-200 family regulates epithelial-mesenchymal transition.* Cancer Res, 2008. **68**(19): p. 7846-54.
38. Tian, X.J., H. Zhang, and J. Xing, *Coupled reversible and irreversible bistable switches underlying TGFβ-induced epithelial to mesenchymal transition.* Biophys J, 2013. **105**(4): p. 1079-89.
39. Zhang, J., et al., *TGF-β-induced epithelial-to-mesenchymal transition proceeds through stepwise activation of multiple feedback loops.* Sci Signal, 2014. **7**(345): p. ra91-ra91.
40. Tian, X.J., et al., *Reciprocal regulation between mRNA and microRNA enables a bistable switch that directs cell fate decisions.* FEBS Lett, 2016. **590**(19): p. 3443-3455.
41. Lu, M., et al., *MicroRNA-based regulation of epithelial-hybrid-mesenchymal fate determination.* Proc Natl Acad Sci U S A, 2013. **110**(45): p. 18144-9.
42. Hong, T., et al., *An Ovol2-Zeb1 Mutual Inhibitory Circuit Governs Bidirectional and Multi-step Transition between Epithelial and Mesenchymal States.* PLoS Comput Biol, 2015. **11**(11): p. e1004569.
43. Jolly, M.K., et al., *Stability of the hybrid epithelial/mesenchymal phenotype.* Oncotarget, 2016. **7**(19): p. 27067-84.
44. Jia, D., et al., *OVOL guides the epithelial-hybrid-mesenchymal transition.* Oncotarget, 2015. **6**(17): p. 15436-48.
45. Steinway, S.N., et al., *Network Modeling of TGFβ Signaling in Hepatocellular Carcinoma Epithelial-to-Mesenchymal Transition Reveals Joint Sonic Hedgehog and Wnt Pathway Activation.* Cancer Research, 2014. **74**(21): p. 5963-5977.

46. Steinway, S.N., et al., *Combinatorial interventions inhibit TGFβ-driven epithelial-to-mesenchymal transition and support hybrid cellular phenotypes.* Npj Systems Biology And Applications, 2015. **1**: p. 15014.
47. Font-Clos, F., S. Zapperi, and C.A.M. La Porta, *Topography of epithelial-mesenchymal plasticity.* Proc Natl Acad Sci U S A, 2018. **115**(23): p. 5902-5907.
48. Tan, T.Z., et al., *Epithelial-mesenchymal transition spectrum quantification and its efficacy in deciphering survival and drug responses of cancer patients.* EMBO Mol Med, 2014. **11**(10): p. 201404208.
49. Huang, R.Y., et al., *An EMT spectrum defines an anoikis-resistant and spheroidogenic intermediate mesenchymal state that is sensitive to e-cadherin restoration by a src-kinase inhibitor, saracatinib (AZD0530).* Cell Death Dis, 2013. **4**: p. e915.
50. Tan, T.Z., et al., *Functional genomics identifies five distinct molecular subtypes with clinical relevance and pathways for growth control in epithelial ovarian cancer.* EMBO Mol Med, 2013. **5**(7): p. 983-98.
51. Schliekelman, M.J., et al., *Molecular portraits of epithelial, mesenchymal, and hybrid States in lung adenocarcinoma and their relevance to survival.* Cancer Res, 2015. **75**(9): p. 1789-800.
52. Pastushenko, I., et al., *Identification of the tumour transition states occurring during EMT.* Nature, 2018.
53. LeBleu, V.S., et al., *Origin and function of myofibroblasts in kidney fibrosis.* Nat Med, 2013. **19**(8): p. 1047-1053.
54. Leung, K.C.W., M. Tonelli, and M.T. James, *Chronic kidney disease following acute kidney injury[mdash]risk and outcomes.* Nat Rev Nephrol, 2013. **9**(2): p. 77-85.
55. Kalluri, R. and E.G. Neilson, *Epithelial-mesenchymal transition and its implications for fibrosis.* J Clin Invest, 2003. **112**(12): p. 1776-84.
56. Stone, R.C., et al., *Epithelial-mesenchymal transition in tissue repair and fibrosis.* Cell Tissue Res, 2016. **365**(3): p. 495-506.
57. Li, M., et al., *Epithelial-mesenchymal transition: An emerging target in tissue fibrosis.* Experimental biology and medicine (Maywood, N.J.), 2016. **241**(1): p. 1-13.
58. Lovisa, S., et al., *Epithelial-to-mesenchymal transition induces cell cycle arrest and parenchymal damage in renal fibrosis.* Nat Med, 2015. **21**(9): p. 998-1009.
59. Grande, M.T., et al., *Snail1-induced partial epithelial-to-mesenchymal transition drives renal fibrosis in mice and can be targeted to reverse established disease.* Nat Med, 2015. **21**(9): p. 989-97.
60. Tian, X.-J., et al., *Trade-off between Quick Repair and Complete Resolution Complicates Recovery from Kidney Injury.* arXiv, 2018: p. 1810.03687v1.
61. Clevers, H. and R. Nusse, *Wnt/β-Catenin Signaling and Disease.* Cell, 2012. **149**(6): p. 1192-1205.
62. Farin, H.F., et al., *Visualization of a short-range Wnt gradient in the intestinal stem-cell niche.* Nature, 2016. **530**(7590): p. 340-343.
63. Perrimon, N., C. Pitsouli, and B.-Z. Shilo, *Signaling Mechanisms Controlling Cell Fate and Embryonic Patterning.* Cold Spring Harbor Perspectives in Biology, 2012. **4**(8).
64. Hrabe, J., S. Hrabetova, and K. Segeth, *A model of effective diffusion and tortuosity in the extracellular space of the brain.* Biophys J, 2004. **87**(3): p. 1606-17.
65. Nicholson, C. and S. Hrabetova, *Brain Extracellular Space: The Final Frontier of Neuroscience.* Biophys J, 2017.
66. Chen, K.C. and C. Nicholson, *Changes in brain cell shape create residual extracellular space volume and explain tortuosity behavior during osmotic challenge.* Proc. Natl. Acad. Sci. U.S.A, 2000. **97**(15): p. 8306-8311.

67. Syková, E. and C. Nicholson, *Diffusion in Brain Extracellular Space.* Physiological Reviews, 2008. **88**(4): p. 1277.
68. Tao, A., L. Tao, and C. Nicholson, *Cell cavities increase tortuosity in brain extracellular space.* J. Theor. Biol 2005. **234**(4): p. 525-536.
69. Tao, L. and C. Nicholson, *Maximum geometrical hindrance to diffusion in brain extracellular space surrounding uniformly spaced convex cells.* J. Theor. Biol , 2004. **229**(1): p. 59-68.
70. Xiao, F., J. Hrabe, and S. Hrabetova, *Anomalous Extracellular Diffusion in Rat Cerebellum.* Biophys J, 2015. **108**(9): p. 2384-2395.
71. Gelens, L., G.A. Anderson, and J.E. Ferrell, Jr., *Spatial trigger waves: positive feedback gets you a long way.* Molecular biology of the cell, 2014. **25**(22): p. 3486-3493.
72. Zhang, J., et al., *Pathway crosstalk enables cells to interpret TGF-β duration.* npj Systems Biology and Applications, 2018. **4**(1): p. 18.
73. Cheong, R., et al., *Information transduction capacity of noisy biochemical signaling networks.* Science, 2011. **334**(6054): p. 354-8.
74. Selimkhanov, J., et al., *Accurate information transmission through dynamic biochemical signaling networks.* Science, 2014. **346**(6215): p. 1370-1373.
75. Kaufhold, S. and B. Bonavida, *Central role of Snail1 in the regulation of EMT and resistance in cancer: a target for therapeutic intervention.* J Exp Clin Cancer Res, 2014. **33**: p. 62.
76. Vega, S., et al., *Snail blocks the cell cycle and confers resistance to cell death.* Genes Dev, 2004. **18**(10): p. 1131-43.
77. Varga, J. and F.R. Greten, *Cell plasticity in epithelial homeostasis and tumorigenesis.* Nature cell biology, 2017. **19**(10): p. 1133.
78. Grosse-Wilde, A., et al., *Stemness of the hybrid epithelial/mesenchymal state in breast cancer and its association with poor survival.* PLoS one, 2015. **10**(5): p. e0126522.
79. Jolly, M.K., et al., *Implications of the Hybrid Epithelial/Mesenchymal Phenotype in Metastasis.* Frontiers in Oncology, 2015. **5**(155).
80. MacLean, A.L., T. Hong, and Q. Nie, *Exploring intermediate cell states through the lens of single cells.* Current Opinion in Systems Biology, 2018.
81. Ta, C.H., Q. Nie, and T. Hong, *Controlling stochasticity in epithelial-mesenchymal transition through multiple intermediate cellular states.* Discrete and continuous dynamical systems. Series B, 2016. **21**(7): p. 2275.
82. Goossens, S., et al., *EMT transcription factors in cancer development re-evaluated: Beyond EMT and MET.* Biochimica et Biophysica Acta (BBA)-Reviews on Cancer, 2017.
83. Shibue, T. and R.A. Weinberg, *EMT, CSCs, and drug resistance: the mechanistic link and clinical implications.* Nature reviews Clinical oncology, 2017. **14**(10): p. 611.
84. Zhang, J., X.-J. Tian, and J. Xing, *Signal transduction pathways of EMT induced by TGF-β, SHH, and WNT and their crosstalks.* Journal of clinical medicine, 2016. **5**(4): p. 41.
85. Yang, L., et al., *Epithelial cell cycle arrest in G2/M mediates kidney fibrosis after injury.* Nat Med, 2010. **16**(5): p. 535-543.
86. Yang, Y., et al., *Transforming growth factor-[beta]1 induces epithelial-to-mesenchymal transition and apoptosis via a cell cycle-dependent mechanism.* Oncogene, 2006. **25**(55): p. 7235-7244.
87. Iordanskaia, T. and A. Nawshad, *Mechanisms of transforming growth factor beta induced cell cycle arrest in palate development.* J Cell Physiol, 2011. **226**(5): p. 1415-24.
88. Kwon, J.S., et al., *Controlling Depth of Cellular Quiescence by an Rb-E2F Network Switch.* Cell Reports, 2017. **20**(13): p. 3223-3235.
89. Tran, D.D., et al., *Temporal and spatial cooperation of Snail1 and Twist1 during epithelial-mesenchymal transition predicts for human breast cancer recurrence.* Mol Cancer Res, 2011. **9**(12): p. 1644-57.

90. Li, Q.-Q., et al., *Twist1-Mediated Adriamycin-Induced Epithelial-Mesenchymal Transition Relates to Multidrug Resistance and Invasive Potential in Breast Cancer Cells.* Clinical Cancer Research, 2009. **15**(8): p. 2657-2665.
91. Barrallo-Gimeno, A. and M.A. Nieto, *The Snail genes as inducers of cell movement and survival: implications in development and cancer.* Development, 2005. **132**(14): p. 3151-3161.
92. Franco, D.L., et al., *Snail1 suppresses TGF-β-induced apoptosis and is sufficient to trigger EMT in hepatocytes.* Journal of Cell Science, 2010. **123**(20): p. 3467-3477.
93. Friedl, P. and K. Wolf, *Tumour-cell invasion and migration: diversity and escape mechanisms.* Nature Reviews Cancer, 2003. **3**: p. 362.
94. David, C.J., et al., *TGF-beta Tumor Suppression through a Lethal EMT.* Cell, 2016. **164**(5): p. 1015-30.
95. Ding, A.X., et al., *CasExpress reveals widespread and diverse patterns of cell survival of caspase-3 activation during development in vivo.* eLife, 2016. **5**: p. e10936.
96. Sun, G., et al., *A molecular signature for anastasis, recovery from the brink of apoptotic cell death.* J Cell Biol., 2017.
97. Tang, H.L., et al., *Cell survival, DNA damage, and oncogenic transformation after a transient and reversible apoptotic response.* Molecular Biology of the Cell, 2012. **23**(12): p. 2240-2252.
98. Liu, X. and D. Fan, *The epithelial-mesenchymal transition and cancer stem cells: functional and mechanistic links.* Curr Pharm Des, 2015. **21**(10): p. 1279-91.
99. Scheel, C. and R.A. Weinberg, *Cancer stem cells and epithelial-mesenchymal transition: concepts and molecular links.* Semin Cancer Biol, 2012. **22**(5-6): p. 396-403.
100. Morel, A.P., et al., *Generation of breast cancer stem cells through epithelial-mesenchymal transition.* PLoS One, 2008. **3**(8): p. e2888.
101. Singh, A. and J. Settleman, *EMT, cancer stem cells and drug resistance: an emerging axis of evil in the war on cancer.* Oncogene, 2010. **29**(34): p. 4741-4751.
102. Campbell, L.L. and K. Polyak, *Breast tumor heterogeneity: cancer stem cells or clonal evolution?* Cell Cycle, 2007. **6**(19): p. 2332-8.
103. Boman, B.M. and M.S. Wicha, *Cancer stem cells: a step toward the cure.* J Clin Oncol, 2008. **26**(17): p. 2795-9.
104. Gupta, P.B., C.L. Chaffer, and R.A. Weinberg, *Cancer stem cells: mirage or reality?* Nat Med, 2009. **15**(9): p. 1010-2.
105. Medema, J.P., *Cancer stem cells: the challenges ahead.* Nat Cell Biol, 2013. **15**(4): p. 338-44.
106. Reya, T., et al., *Stem cells, cancer, and cancer stem cells.* Nature, 2001. **414**(6859): p. 105-11.
107. Gupta, P.B., et al., *Stochastic state transitions give rise to phenotypic equilibrium in populations of cancer cells.* Cell, 2011. **146**(4): p. 633-44.
108. Friedmann-Morvinski, D. and I.M. Verma, *Dedifferentiation and reprogramming: origins of cancer stem cells.* EMBO reports, 2014. **15**(3): p. 244-253.
109. Gupta, P.B., C.L. Chaffer, and R.A. Weinberg, *Cancer stem cells: mirage or reality?* Nature medicine, 2009. **15**(9): p. 1010.
110. Mani, S.A., et al., *The epithelial-mesenchymal transition generates cells with properties of stem cells.* Cell, 2008. **133**(4): p. 704-15.
111. Morel, A.-P., et al., *Generation of Breast Cancer Stem Cells through Epithelial-Mesenchymal Transition.* PLoS ONE, 2008. **3**(8): p. e2888.
112. McDonald, O.G., et al., *Genome-scale epigenetic reprogramming during epithelial-to-mesenchymal transition.* Nat Struct Mol Biol, 2011. **18**(8): p. 867-74.
113. Fu, J., et al., *The TWIST/Mi2/NuRD protein complex and its essential role in cancer metastasis.* Cell Research, 2010. **21**: p. 275.

114. Lin, Y., et al., *The SNAG domain of Snail1 functions as a molecular hook for recruiting lysine‐specific demethylase 1.* The EMBO journal, 2010. **29**(11): p. 1803-1816.
115. Lin, T., et al., *Requirement of the histone demethylase LSD1 in Snai1-mediated transcriptional repression during epithelial-mesenchymal transition.* Oncogene, 2010. **29**: p. 4896.
116. Chaffer, Christine L., et al., *Poised Chromatin at the ZEB1 Promoter Enables Breast Cancer Cell Plasticity and Enhances Tumorigenicity.* Cell, 2013. **154**(1): p. 61-74.
117. Zhang, J., et al., *Spatial clustering and common regulatory elements correlate with TGF-β induced concerted gene expression. PLoS. Comp. Biol.* 2018. **Accepted**.
118. Tian, X.J., et al., *Achieving diverse and monoallelic olfactory receptor selection through dual-objective optimization design.* Proc Natl Acad Sci U S A, 2016. **113**(21): p. E2889-98.
119. Zhang, H., et al., *Statistical Mechanics Model for the Dynamics of Collective Epigenetic Histone Modification.* Phys. Rev. Lett., 2014. **112**(6): p. 068101.
120. Dodd, I.B., et al., *Theoretical analysis of epigenetic cell memory by nucleosome modification.* Cell, 2007. **129**(4): p. 813-22.
121. Angel, A., et al., *A Polycomb-based switch underlying quantitative epigenetic memory.* Nature, 2011. **476**(7358): p. 105-108.
122. Cortini, R., et al., *The physics of epigenetics.* Reviews of Modern Physics, 2016. **88**(2): p. 025002.
123. Michieletto, D., E. Orlandini, and D. Marenduzzo, *Polymer model with Epigenetic Recoloring Reveals a Pathway for the \textit{de novo} Establishment and 3D Organization of Chromatin Domains.* Physical Review X, 2016. **6**(4): p. 041047.
124. Beerling, E., et al., *Plasticity between epithelial and mesenchymal states unlinks EMT from metastasis-enhancing stem cell capacity.* Cell reports, 2016. **14**(10): p. 2281-2288.
125. Mandal, M., et al., *Modeling continuum of epithelial mesenchymal transition plasticity.* Integr Biol (Camb), 2016. **8**(2): p. 167-76.
126. Dong, J., et al., *Single-cell RNA-seq analysis unveils a prevalent epithelial/mesenchymal hybrid state during mouse organogenesis.* Genome Biol, 2018. **19**(1): p. 31.
127. Puram, S.V., et al., *Single-Cell Transcriptomic Analysis of Primary and Metastatic Tumor Ecosystems in Head and Neck Cancer.* Cell, 2017. **171**(7): p. 1611-1624.e24.
128. Puram, S.V., A.S. Parikh, and I. Tirosh, *Single cell RNA-seq highlights a role for a partial EMT in head and neck cancer.* Mol Cell Oncol, 2018. **5**(3): p. e1448244.
129. Chung, W., et al., *Single-cell RNA-seq enables comprehensive tumour and immune cell profiling in primary breast cancer.* Nat Commun, 2017. **8**: p. 15081.
130. Kuo, C.S. and M.A. Krasnow, *Formation of a Neurosensory Organ by Epithelial Cell Slithering.* Cell, 2015. **163**(2): p. 394-405.
131. Burute, M., et al., *Polarity Reversal by Centrosome Repositioning Primes Cell Scattering during Epithelial-to-Mesenchymal Transition.* Developmental Cell, 2017. **40**(2): p. 168-184.
132. Kherlopian, A.R., et al., *A review of imaging techniques for systems biology.* BMC systems biology, 2008. **2**(1): p. 74.
133. Uchida, S., *Image processing and recognition for biological images.* Development, growth & differentiation, 2013. **55**(4): p. 523-549.
134. Roeder, A.H., et al., *A computational image analysis glossary for biologists.* Development, 2012. **139**(17): p. 3071-3080.
135. Rueden, C.T., et al., *ImageJ2: ImageJ for the next generation of scientific image data.* BMC Bioinformatics, 2017. **18**(1): p. 529.
136. Carpenter, A.E., et al., *CellProfiler: image analysis software for identifying and quantifying cell phenotypes.* Genome Biol, 2006. **7**(10): p. R100.
137. LeCun, Y., Y. Bengio, and G. Hinton, *Deep learning.* Nature, 2015. **521**(7553): p. 436-444.

138. Van Valen, D.A., et al., *Deep Learning Automates the Quantitative Analysis of Individual Cells in Live-Cell Imaging Experiments.* PLoS Comput Biol, 2016. **12**(11): p. e1005177.
139. Wang, W., et al., *Learn to segment single cells with deep distance estimator and deep cell detector.* arXiv preprint arXiv:1803.10829, 2018.
140. Eschweiler, D., et al., *CNN-based Preprocessing to Optimize Watershed-based Cell Segmentation in 3D Confocal Microscopy Images.* arXiv preprint arXiv:1810.06933, 2018.
141. Chen, Y.-J., et al., *Rapid, modular, and cost-effective generation of donor DNA constructs for CRISPR-based gene knock-in* bioRxiv, 2017.
142. Li, H., et al., *Design and specificity of long ssDNA donors for CRISPR-based knock-in.* bioRxiv, 2017.
143. Bertrand, E., et al., *Localization of ASH1 mRNA Particles in Living Yeast.* Mol Cell, 1998. **2**(4): p. 437-445.
144. Chen, B., et al., *Dynamic Imaging of Genomic Loci in Living Human Cells by an Optimized CRISPR/Cas System.* Cell, 2013. **155**(7): p. 1479-1491.
145. Chen, B., J. Guan, and B. Huang, *Imaging Specific Genomic DNA in Living Cells.* Annual Review of Biophysics, 2016. **45**(1): p. 1-23.
146. Guan, J., et al., *Tracking Multiple Genomic Elements Using Correlative CRISPR Imaging and Sequential DNA FISH.* Biophysical Journal, 2017. **112**(6): p. 1077-1084.
147. Wang, S., et al., *An RNA-aptamer-based two-color CRISPR labeling system.* Scientific Reports, 2016. **6**: p. 26857.
148. Kang, H., et al., *Confinement-Induced Glassy Dynamics in a Model for Chromosome Organization.* Physical Review Letters, 2015. **115**(19): p. 198102.
149. Ma, H., et al., *Multiplexed labeling of genomic loci with dCas9 and engineered sgRNAs using CRISPRainbow.* Nat Biotech, 2016. **34**(5): p. 528-530.
150. Takei, Y., et al., *Multiplexed Dynamic Imaging of Genomic Loci by Combined CRISPR Imaging and DNA Sequential FISH.* Biophysical Journal, 2017. **112**(9): p. 1773-1776.
151. Shao, S., et al., *Illuminating the structure and dynamics of chromatin by fluorescence labeling.* Frontiers in Biology, 2017.
152. Shao, S., et al., *Multiplexed sgRNA Expression Allows Versatile Single Non-repetitive DNA Labeling and Endogenous Gene Regulation.* ACS Synth Biol, 2018. **7**(1): p. 176-186.
153. Ma, T., et al., *Developing novel methods to image and visualize 3D genomes.* Cell Biology and Toxicology, 2018.
154. Chen, B., et al., *Expanding the CRISPR imaging toolset with Staphylococcus aureus Cas9 for simultaneous imaging of multiple genomic loci.* Nucleic Acids Research, 2016.
155. Fu, Y., et al., *CRISPR-dCas9 and sgRNA scaffolds enable dual-colour live imaging of satellite sequences and repeat-enriched individual loci.* Nat Commun, 2016. **7**: p. 11707.
156. Qin, P., et al., *Live cell imaging of low- and non-repetitive chromosome loci using CRISPR-Cas9.* Nat Commun, 2017. **8**: p. 14725.
157. Gu, B., et al., *Transcription-coupled changes in nuclear mobility of mammalian cis-regulatory elements.* Science, 2018. **359**(6379): p. 1050.
158. Maass, P.G., et al., *Inter-chromosomal Contact Properties in Live-Cell Imaging and in Hi-C.* Mol Cell, 2018. **69**(6): p. 1039-1045.e3.
159. Liao, H.-K., et al., *In vivo Target Gene Activation via CRISPR/Cas9 Trans-Mediated epigenetic Modulation.* Cell, 2017. **171**(7): p. 1495-1507.
160. Qi, L.S., et al., *Repurposing CRISPR as an RNA-guided platform for sequence-specific control of gene expression.* Cell, 2013. **152**(5): p. 1173-83.
161. Hilton, I.B., et al., *Epigenome editing by a CRISPR-Cas9-based acetyltransferase activates genes from promoters and enhancers.* Nat Biotech, 2015. **33**(5): p. 510-517.

162. Lim, Y.-Y., et al., *Epigenetic modulation of the miR-200 family is associated with transition to a breast cancer stem-cell-like state.* Journal of Cell Science, 2013. **126**(10): p. 2256-2266.
163. Davalos, V., et al., *Dynamic epigenetic regulation of the microRNA-200 family mediates epithelial and mesenchymal transitions in human tumorigenesis.* Oncogene, 2012. **31**(16): p. 2062-2074.
164. Brangwynne, C.P., *Phase transitions and size scaling of membrane-less organelles.* J Cell Biol, 2013. **203**(6): p. 875.
165. Steinway, S.N., et al., *Network modeling of TGFβ signaling in hepatocellular carcinoma epithelial-to-mesenchymal transition reveals joint Sonic hedgehog and Wnt pathway activation.* Cancer research, 2014. **74**(21): p. 5963-5977.
166. Xing, J., *Mapping between dissipative and Hamiltonian systems.* J. Phys. A: Math. Theor., 2010. **43**: p. 375003.
167. Hilfinger, A., et al., *Constraints on Fluctuations in Sparsely Characterized Biological Systems.* Phys Rev Lett, 2016. **116**(5): p. 058101.
168. Hilfinger, A., Thomas M. Norman, and J. Paulsson, *Exploiting Natural Fluctuations to Identify Kinetic Mechanisms in Sparsely Characterized Systems.* Cell Systems, 2016. **2**(4): p. 251-259.

*Legend:*

**Figure 1**. **Example mathematical analysis of EMT regulation.** (A) EMT proceeds through partial EMT states. (B) The core regulatory network regulates TGF-β-induced EMT. (C) Bifurcation analysis of the mathematical model predicts a Cascading Bistable Switches (CBS) mechanism with three EMT phenotypes: Epithelial state, partial EMT state (mixed feature of epithelial and mesenchymal states), and Mesenchymal state. E-cadherin and Vimentin are epithelial and mesenchymal markers, respectively. Pointed and blunt-ended arrows represent activation and inhibition, respectively. Adapted from [39].

**Figure 2 Single cell flow cytometry data of human MCF10A cells treated with different concentrations of TGF-β supports the CBS model.** E, P, and M refer to the epithelial, pEMT, and mesenchymal states, respectively. Adapted from [39].

**Figure 3 Cells commit to EMT only after receiving stimuli with strength and duration above certain threshold values.** (A) Model prediction that TGF-β with different duration and strength induces different EMT states. Adapted from [38]. (B) Schematic of the pathway crosstalk model revealed through combined modeling and quantitative experiments.

**Figure 4 Two possible coupling mechanisms between EMT and cell cycle arrest.** (A) Scheme1 predicts four possible cell states, Epithelial, partial EMT-α (Pα), partial EMT-β (Pβ), Mesenchymal. (B) Scheme 2 predicts three cell states with only one partial EMT state (Pα) associated with cell cycle arrest. It is noted that Pα is under cell cycle arrest and thus non-proliferative while Pβ is proliferative. The stability or the depth of the partial EMT state Pα depends on the level of P21.

**Figure 5 Human HK2 cells treated with TGF-β undergo different cell fates as reflected by morphological change.** Cell shapes were analyzed from time-lapse movies. Unpublished data from the Xing lab.

**Figure 6 Possible state networks of the EMT-CSC system.** (A) Complete state network with weak coupling between the two programs. (B) Partially populated state network with strong coupling, so only some transition pathways dominate. (C) Alternative possible partially

populated state network with dominated pathways different from panel B. Solid arrows represent the dominating pathways.

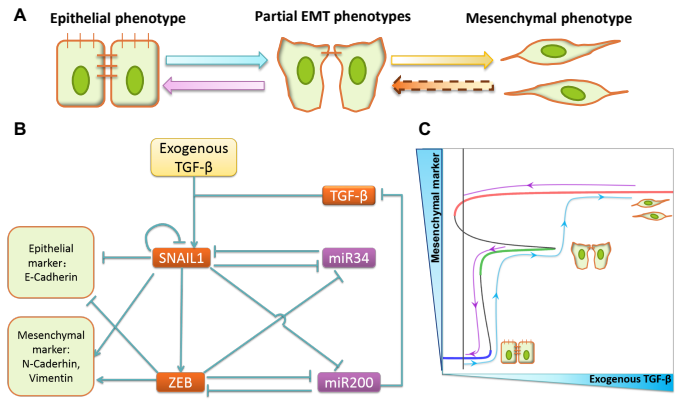

Figure 1

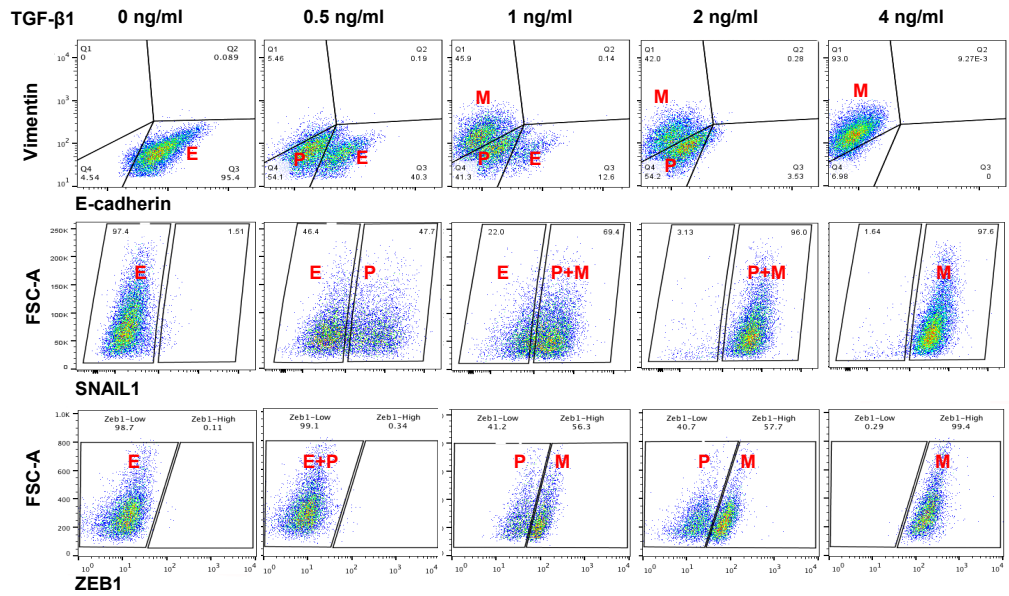

Figure 2

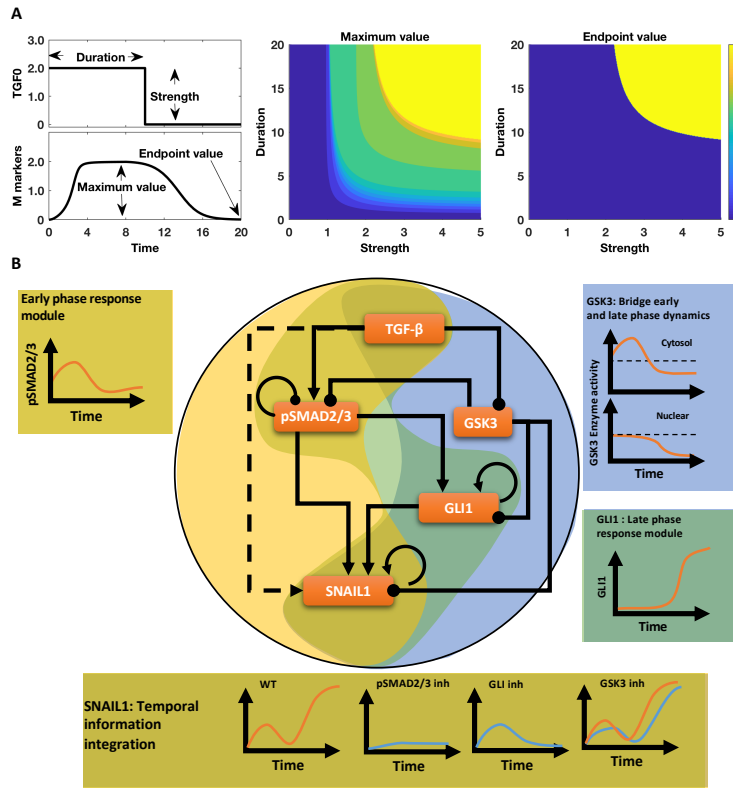

Figure 3

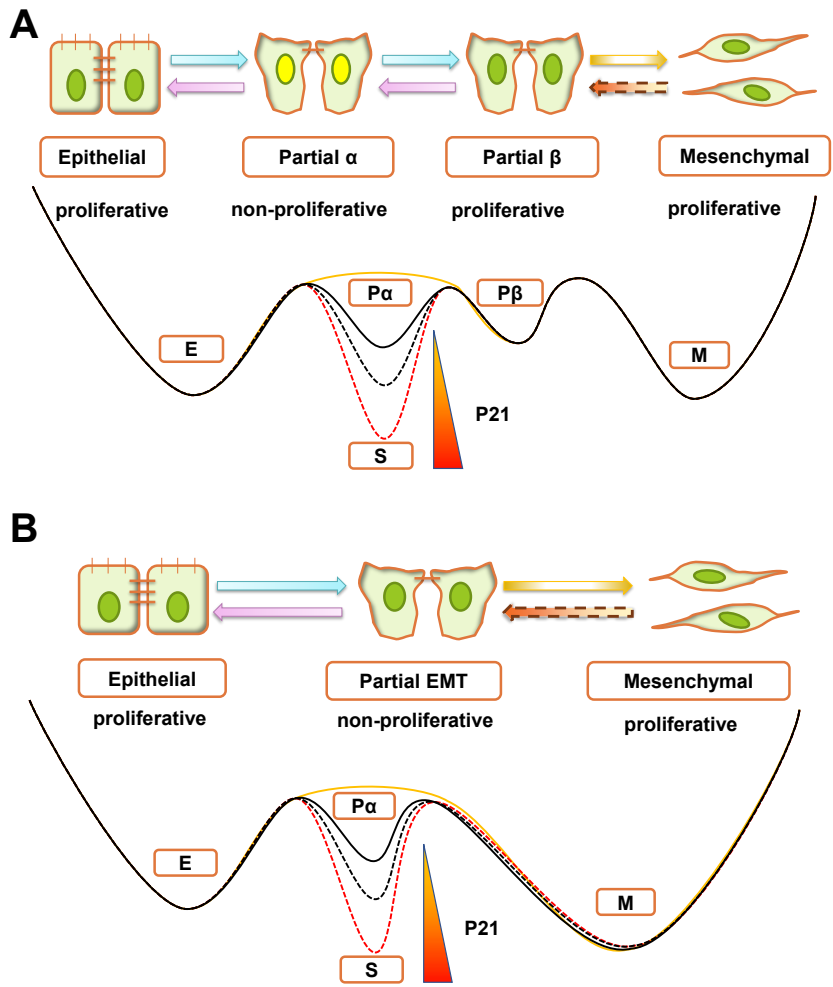

Figure 4

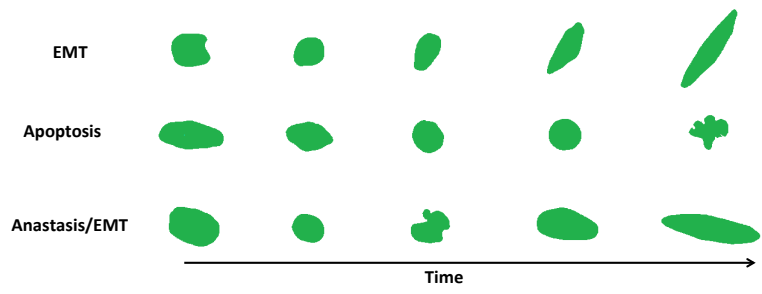

Figure 5

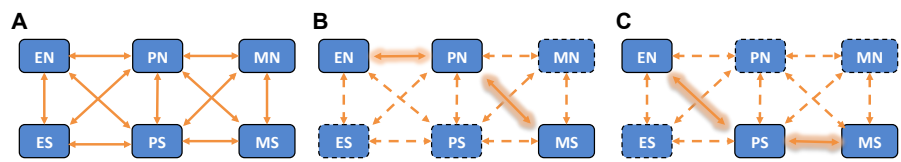

Figure 6